\documentclass[12pt,a4paper]{article}
\usepackage{amsmath, amssymb, amsthm}
\usepackage{graphicx,amssymb,amsfonts,amsbsy,amsmath,amsthm}
\usepackage{mathtools}
\usepackage[utf8]{inputenc}
\usepackage{floatrow}
\usepackage{float}
\usepackage{amssymb}
\usepackage[T1]{fontenc}
\usepackage[utf8]{inputenc}
\usepackage{amsthm}
\usepackage{epstopdf}
\usepackage{cite}
\usepackage{anysize}
\usepackage{color,soul}
\usepackage{graphicx}
\usepackage{tabularx}
\usepackage{multirow}
\usepackage{hyperref}
\hypersetup{colorlinks,linkcolor={blue},citecolor={blue},urlcolor={black}}
\usepackage{enumerate}
\marginsize{2cm}{2cm}{1cm}{1cm}

\usepackage{lineno}
\usepackage[english]{babel}
\begin{document}

\title{Heterogeneous population and its resilience to misinformation in vaccination uptake: A dual ODE and network approach}
\date{}
\author{Komal Tanwar$^{1}$\thanks{Both authors contributed equally.}, Viney Kumar$^{2}$\footnotemark[1], Jai Prakash Tripathi$^{1}$\thanks{Corresponding author: jtripathi85@gmail.com}}

\maketitle

\begin{enumerate}
 \item Department of Mathematics, Central University of Rajasthan, Bandarsindri, Kishangarh, Ajmer  305801, Rajasthan, India
 \item Disease modelling lab, Department of Mathematics, School of Natural Sciences, Shiv Nadar Institution of Eminence (SNIoE), NH-91, Gautam Buddha Nagar, Uttar Pradesh, 201314, India
\end{enumerate}
\begin{abstract}
Misinformation about vaccination poses a significant public health threat by reducing vaccination rates and increasing disease burden. Understanding population heterogeneity can aid in recognizing and mitigating the effects of such misinformation, mainly when vaccine effectiveness is low. Our research quantifies the impact of misinformation on vaccination uptake and explores its effects in heterogeneous versus homogeneous populations. We employed a dual approach combining ordinary differential equations (ODE) and complex network models to analyze how different epidemiological parameters influence disease spread and vaccination behaviour. Our results indicate that misinformation significantly lowers vaccination rates, particularly in homogeneous populations, while heterogeneous populations demonstrate greater resilience. Among network topologies, small-world networks achieve higher vaccination rates under varying vaccine efficacies, whereas scale-free networks experience reduced vaccine coverage with higher misinformation amplification.
Notably, cumulative infection remains independent of the disease transmission rate when the vaccine is partially effective. In small-world networks, cumulative infection shows high stochasticity across vaccination rates and misinformation parameters, while cumulative vaccination is highest with higher vaccination rates and lower misinformation coefficients. Public health efforts may have prioritized addressing misinformation to control disease spread, particularly in homogeneous populations and scale-free networks, where its impact is more severe. Building resilience by fostering diverse community networks and promoting reliable vaccine information can boost vaccination rates. Additionally, focusing public health campaigns on small-world networks may result in higher vaccine uptake, even when vaccine efficacy varies. These insights can help public health policymakers in designing effective vaccination strategies that consider population heterogeneity.
\end{abstract}
\section{Introduction}
Preventive measures play a critical role in controlling the spread of infectious diseases. These include practices like hand hygiene, social distancing, and wearing masks \cite{HJ2005}. Although these measures effectively reduce transmission, vaccination remains the primary method for preventing disease. Vaccines have historically eradicated or significantly reduced the prevalence of many infectious diseases, saving millions of lives worldwide \cite{BB2014}. Despite their proven effectiveness, vaccine hesitancy persists, driven by a range of factors such as distrust in the healthcare system, personal beliefs, and misinformation about vaccine safety. Misinformation, often spread through social media and other platforms, plays a significant role in undermining public trust and increasing skepticism about vaccines \cite{jennings2021lack}. False claims linking vaccines to adverse effects or questioning their necessity exacerbate hesitancy. Additionally, cultural, religious, and socioeconomic factors can fuel distrust and skepticism \cite{upenieks2022trust}. Misinformation is defined as false or misleading information that spreads unintentionally \cite{WL2019}. The World Economic Forum listed ``misinformation" as one of the global risks \cite{WE2013}. Studies indicate that exposure to misinformation, mainly through social media, reduces people's willingness to get vaccinated \cite{LS2021}. There were over $5.16$ $\times$ $10^9$ internet users worldwide, accounting for more than half of the global population. With 59.4 \% of these users active on social networks, the impact of misinformation on behaviour, including vaccination attitudes, was significant \cite{LS2021}.\\

Vaccination campaigns face significant challenges convincing individuals exposed to misinformation. \cite{KS2023, LS2021}. False claims related to the MMR vaccine to autism caused a significant drop in vaccination rates, driven by parental concerns \cite{SR2011}. Researchers estimate that 20-30 \% of YouTube videos about emerging infectious diseases contain inaccurate or misleading information \cite{khatri2020youtube}. A recent study by Stephen R. Neely et al. \cite{NS2022} revealed a significant correlation between exposure to misinformation and decreased vaccination rates. They found that for people exposed to six or more misinformation themes, the vaccination rate plummeted to 52.2 \% \cite{NS2022}. To address this challenge, health authorities encourage individuals to actively share positive information about the benefits of vaccination through various platforms, including social media, local healthcare providers, and community leaders. This approach is a part of broader efforts to curb misinformation and support disease prevention. However, integrating these efforts across different strategic levels highlights the critical need for comprehensive approaches to effectively counter misinformation and promote accurate vaccine information.\\

Numerous computational and mathematical models have been proposed/developed to study the impact of misinformation on vaccination campaigns among populations \cite{tambuscio2015fact,zhang2016dynamics,loomba2021measuring,prieto2021vaccination,pierri2022online,mumtaz2022exploring,chen2022coevolving,sun2023finding}. M. Tambuscio et al. \cite{tambuscio2015fact} developed their model to show the impact of misinformation on the epidemic threshold across homogeneous, heterogeneous, and real networks. Z.K. Zhang et al. \cite{zhang2016dynamics} reviewed advances in information diffusion, examining their functional roles and the optimization strategies employed. Sahil Loomba et al. \cite{loomba2021measuring} conducted a randomized controlled trial in the UK and USA to assess the impact of online misinformation about COVID-19 vaccines on people's vaccination intentions. Francesco Pierri et al. \cite{pierri2022online} found that misbeliefs significantly contribute to vaccine hesitancy. Their research explored the connections between misinformation, beliefs, and human behaviour, revealing a negative correlation between misinformation and vaccination uptake rates. Nabeela Mumtaz et al. \cite{mumtaz2022exploring} investigated two interacting contagion models: one for disease transmission and the other for public perceptions of vaccination. Through sensitivity analysis and loop impact analysis, they examined how misinformation and vaccine confidence influence the spread of infectious diseases. Zejun Sun et al.'s \cite{sun2023finding} novel method was inspired by observed trends in information diffusion and the Matthew effect. Bhattacharyya et al. \cite{kumar2023nonlinear} demonstrated that information weighted against vaccination has a detrimental effect on vaccination rates. Existing literature have investigated the spread and impact of misinformation on complex networks \cite{tambuscio2015fact,zhang2016dynamics,mumtaz2022exploring,chen2022coevolving,sun2023finding} and ordinary differential equation (ODE) models \cite{loomba2021measuring,prieto2021vaccination,pierri2022online}.\\

However, these studies have not conducted a direct comparison to understand how the effects of misinformation differ across heterogeneous and homogeneous populations. Moreover, there has not been a thorough investigation of the effect of vaccine efficacy on individuals' vulnerability to misinformation, especially in more homogeneous populations. As a result, there is a critical research gap in examining how varying levels of population heterogeneity, misinformation rates, and vaccination rates interact across different network structures. This interplay is crucial for designing effective vaccination strategies, particularly during a pandemic. To devise targeted public health policies, it is essential to analyze how misinformation dissemination correlates with network characteristics and vaccine efficacy, affecting vaccination rates across diverse populations.\\

In this research, we quantify the impact of misinformation on vaccination uptake and examine its effects in homogeneous and heterogeneous populations. Our findings reveal that misinformation significantly reduces vaccination rates, especially in homogeneous populations, while heterogeneous populations show greater resilience. Small-world networks achieve higher vaccination rates with varying vaccine effectiveness among network topologies, whereas scale-free networks show lower coverage with increased misinformation. Interestingly, when vaccine efficacy is partial, cumulative infection is independent of disease transmission rates. In small-world networks, cumulative infection is highly stochastic across different vaccination rates and misinformation parameters, while cumulative vaccination peaks with higher vaccination rates and lower misinformation levels. These results offer valuable insights for public health policymakers in developing vaccination strategies that consider population heterogeneity.\\

The article is organized as follows: Section 2 details the model components and their integration; Section 3 explains the numerical simulations of the ODE model, network metrics, and various epidemiological parameters. Section 4 presents the discussion and limitations.

\section{Model framework}
We aim to investigate the impact of vaccination misinformation and varying community dynamics on disease prevalence and vaccine coverage. In our model, interactions with neighbours and their personal disease states shape individual vaccination attitudes. Community members actively counter misinformation as they prioritize vaccination to prevent reinfection. Notably, an individual’s vaccination decision does not influence the choices of their family members.
We proposed our model based on two approaches: one for homogeneous populations (using an ODE approach) and another for heterogeneous populations (using a complex network approach). In the network-based model, disease transmission is influenced by the infection status of neighbouring individuals. 

\begin{figure}[H]
		\centering
		\includegraphics[scale=0.2]{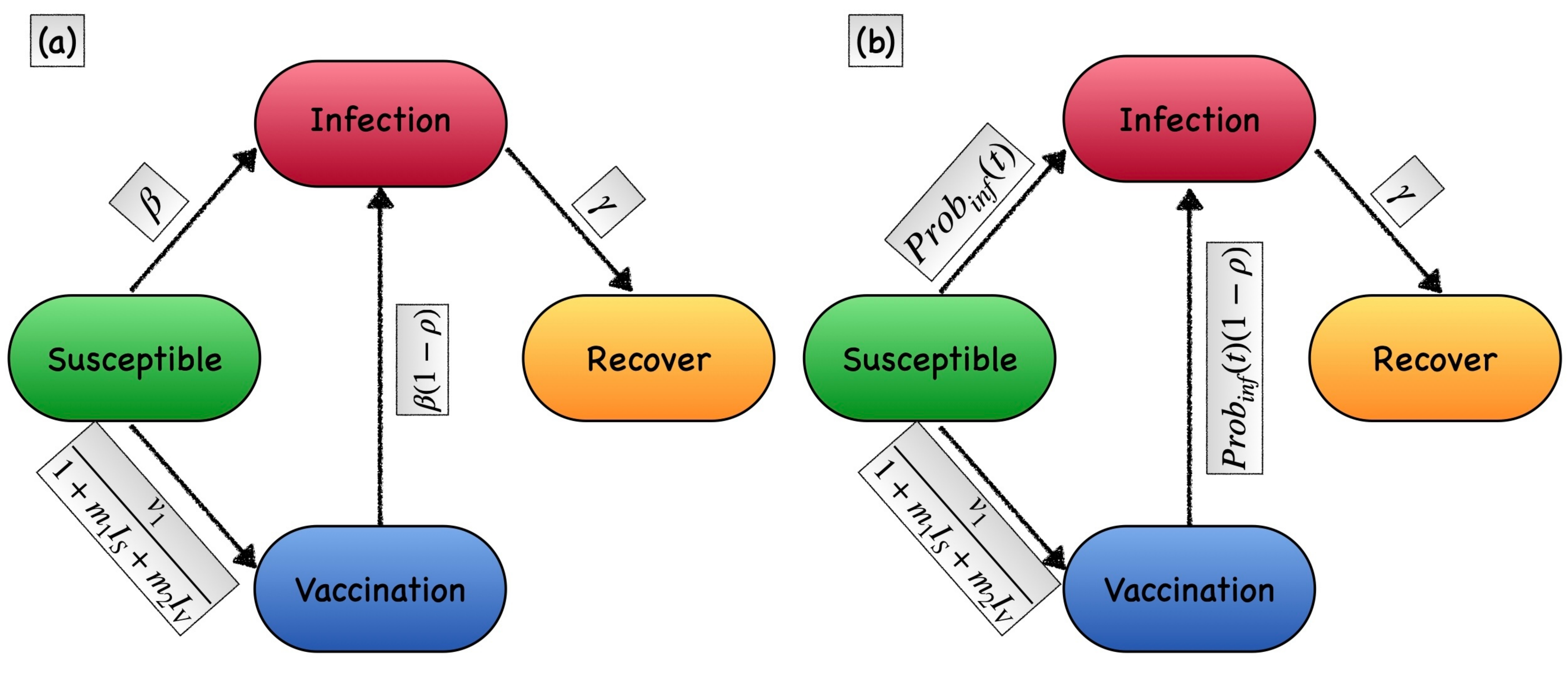}
		\caption{Schematic diagram of the disease model for (a) ODE approach (b) complex-network-based approach}
		\label{Fig:sche}
	\end{figure}

Total population ($N$) is divided into four compartments: susceptible ($S$), infected ($I$), vaccinated ($V$), and recovered ($R$). Here $N = S + I + V + R$. Figure \ref{Fig:sche} presents the schematic of the disease model, and Table \ref{table1} lists the used parameters.
In this model, the susceptible individuals get infected when had contact with infected neighbours and move to the infected class at the rate $\beta$. The infected individuals move to the recovered class after getting recovered from the infection at the rate $\gamma$. As susceptible individuals can contract the infection from their infected neighbours, they may also get a vaccine that provides partial immunity. However, due to the low efficacy of the vaccine (denoted by $\rho$), vaccinated individuals remain susceptible to infection and can still contract the disease from infected individuals. As a result, we define two types of infections: $I_S$ (infection in susceptible individuals) and $I_V$ (infection in vaccinated individuals) \cite{kumar2023nonlinear}, which are described in detail in Table \ref{table1}. Both $I_S$ and $I_V$ contribute to spreading misinformation about the vaccine, governed by parameters $m_1$ and $m_2$, respectively. These parameters affect the overall vaccine uptake in the population. Therefore, we consider the vaccination rate as a function of the number of infected individuals.
	\begin{center}
		$	V(I) =  \dfrac{v_1}{1+(m_1 I_s+m_2 I_v)}.$
	\end{center}
	We chose this function to model the impact of misinformation for the following reasons. In reality, people tend to behave rationally, and when exposed to misinformation about vaccination, they may avoid getting vaccinated to protect themselves. As misinformation spreads, the vaccination rate decreases further. Here we assume that misinformation about vaccines may spread through infected individuals, either directly to susceptible individuals or, if they become infected, after vaccination. We multiply the baseline vaccination rate $v_1$ by the function $\frac{1}{1+(m_1 I_s+m_2 I_v)}$ to reflect the reduced value of vaccination rate due to misinformation spread about vaccines. After the vaccination, we assume that the infected individuals spread more misinformation about vaccines, so we consider the misinformation coefficient $m_2 > m_1$ in our model. The \textbf{ODE-based model} is formulated as follows:\\
\begin{equation}
\begin{aligned}
	\dfrac{dS}{dt} &= -\beta SI -\dfrac{v_1 S}{1+(m_1 I_s+m_2 I_v)}\\
	\dfrac{dV}{dt} &= \dfrac{v_1 S}{1+(m_1 I_s+m_2 I_v)} - \beta (1-\rho) VI\\
	\dfrac{dI}{dt} &= \beta SI +  \beta (1-\rho) VI - \gamma I \\
	\dfrac{dR}{dt} &= \gamma I
	\end{aligned}
	\end{equation}
subject to initial condition:
$S(0) = S_0$, $I(0) = I_0$, $V(0) = V_0$, and $R(0) = R_0$.\\
In addition to the ODE model, we incorporate a complex network approach to capture the diverse interactions among individuals and their influence on disease spread and vaccination behaviours.\\\\
\textbf{Network-based Model framework}

Our network model represents individuals as nodes, with connections symbolizing physical interactions facilitating disease transmission and information exchange.

In this framework, a $j^{th}$ susceptible individual’s risk of infection depends on the disease transmission rate ($\beta$) and the number of infected neighbours ($N_{inf}^{j}$). The probability of getting in contact with infection of a $j^{th}$ susceptible individual at each time step (t) is defined as \cite{kumar2023nonlinear}:

\[ \text{Prob}_{\textit{inf}}(t) = 1 - (1 - \beta)^{N_{inf}^{j}(t)} \]
\begin{figure}[H]
    \centering
    \includegraphics[scale=0.7]{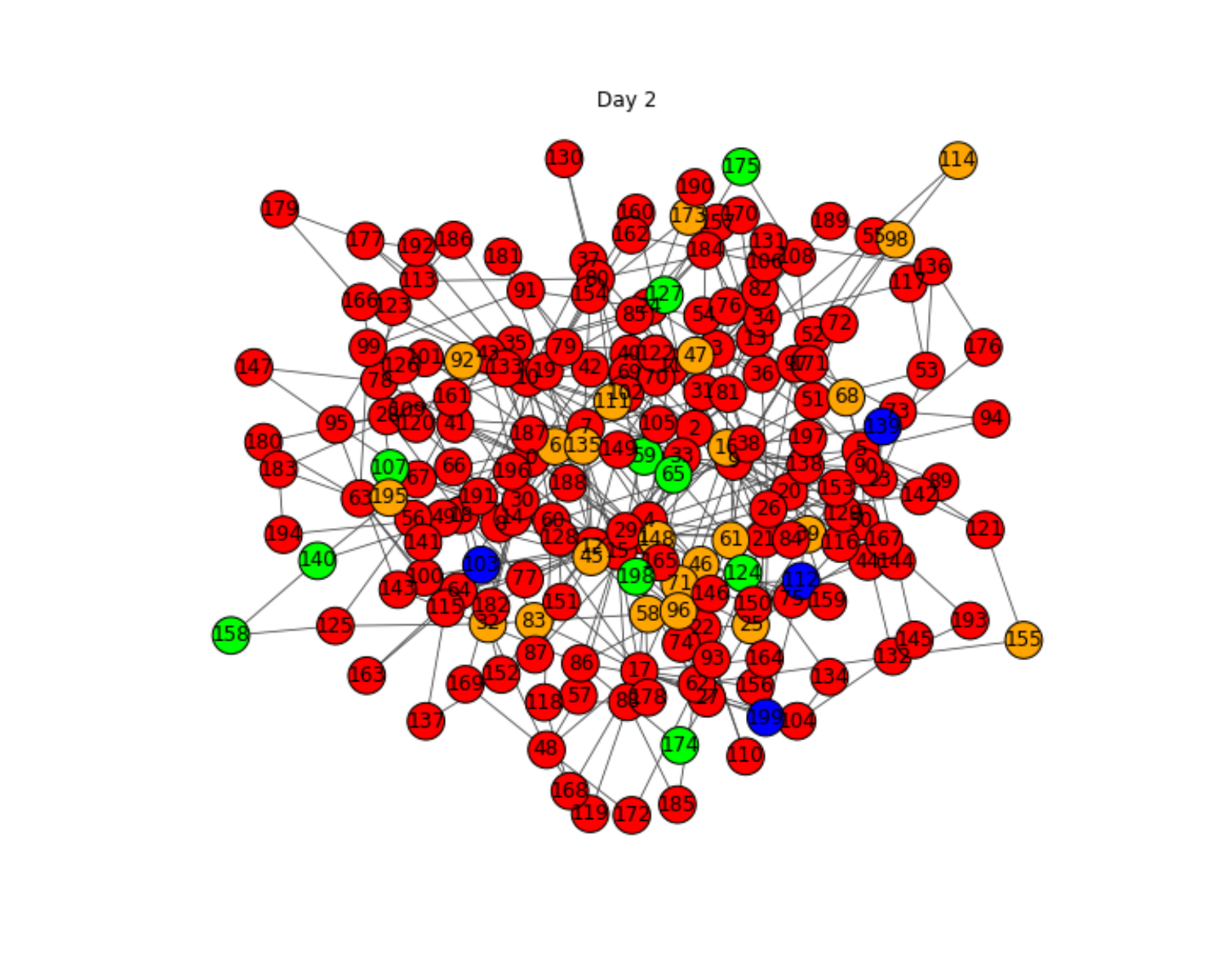}
    \caption{The Figure illustrates the network structure on Day 2. Colours represent different states: green for susceptible, red for infected, blue for vaccinated, and yellow for recovered individuals. The network consists of 200 nodes and 399 edges.}
    \label{Fig:netw_sch}
\end{figure}
An infected individual may recover and become immune with a probability of $\gamma$ per time step. Furthermore, individuals may choose vaccinations based on community-wide vaccination rates. Due to partial vaccine efficacy ($\rho$), vaccinated individuals still face a risk of infection, determined by the probability $Prob_{inf} \times (1-\rho)$ \cite{kumar2023nonlinear} and the infection status of their neighbours.\\

This network-based approach allows us to model the complex dynamics of disease spread and vaccination uptake in heterogeneous populations. While the ODE model assumes each individual has the same number of connections due to homogeneity, the network model captures the variability in the number of neighbours, making it more representative of real-world heterogeneity. Figure \ref{Fig:netw_sch} depicts the disease dynamics within a networked population. By comparing the outcomes of the ODE model and the network approach, we gain deeper insights into how varying population structures influence disease prevalence and vaccination coverage.\\
	
	\begin{table}[H]
		\caption{Description of baseline parameters \& variables}  \label{table1}
		\begin{tabular}{|p{2cm}|p{7cm}|p{4cm}}
			\hline
			Variables & Description  & Values \\
			\hline
            $N$ & total population \\
            $S(t)$ & susceptible population at time $t$\\
            $I(t)$ & total infected population at time $t$\\
             $I_S (t)$ & infected individual who get an infection from susceptible state at time $t$\\
            $I_V (t)$ & infected individual who get an infection from the vaccinated state due to $\rho \neq 1$ at time $t$\\
            $V(t)$ & vaccinated population at time $t$\\
            $R(t)$ & recovered population at time $t$\\
            \hline
            Parameter & Description  &  \\
			\hline
            $\beta$ & disease transmission rate (per day) & 0.667 \cite{kumar2023nonlinear}\\
			$v_1$ & vaccination rate (per day) & [0,1], 0.5 calibrated \\
            $\rho$ & vaccine efficacy & [0,1], 0.8 \cite{kumar2023nonlinear}\\
			$m_1$ & misinformation amplification parameter for $I_S$ state individuals & [0,0.1], 0.05 calibrated\\
			$m_2$ & misinformation amplification parameter for $I_V$ state individuals & [0,0.2], 0.08 calibrated\\
			$1/\gamma$ & infectious period & 7 days \cite{kumar2023nonlinear} \\
			\hline
		\end{tabular}
	\end{table}
 \section{Results}
 \subsection{Numerical Simulations}
 Our model employed a complex network approach, primarily focusing on a scale-free network model \cite{jeong2003measuring}. Concurrently, we conducted simulations using random and small-world network models \cite{dereich2009random,newman2001clustering}. These networks are generated using the igraph-python package, each consisting of 5000 nodes and 9997 edges \cite{batagelj2018python}. The network's mean degree distribution is $\approx$ 2. Supplementary Figures \textbf{S1, S2, and S3} illustrate the degree distributions of these networks. All the complex network numerical simulations were carried out using Python, while MATLAB's built-in \textit{ode45} functions facilitated numerical simulations for the ODE model and generated all plots. We have simulated our model using both approaches and different epidemiological parameters; results are discussed in the following subsections.
 \begin{figure}[H]
	    \centering
	    \includegraphics[scale=0.6]{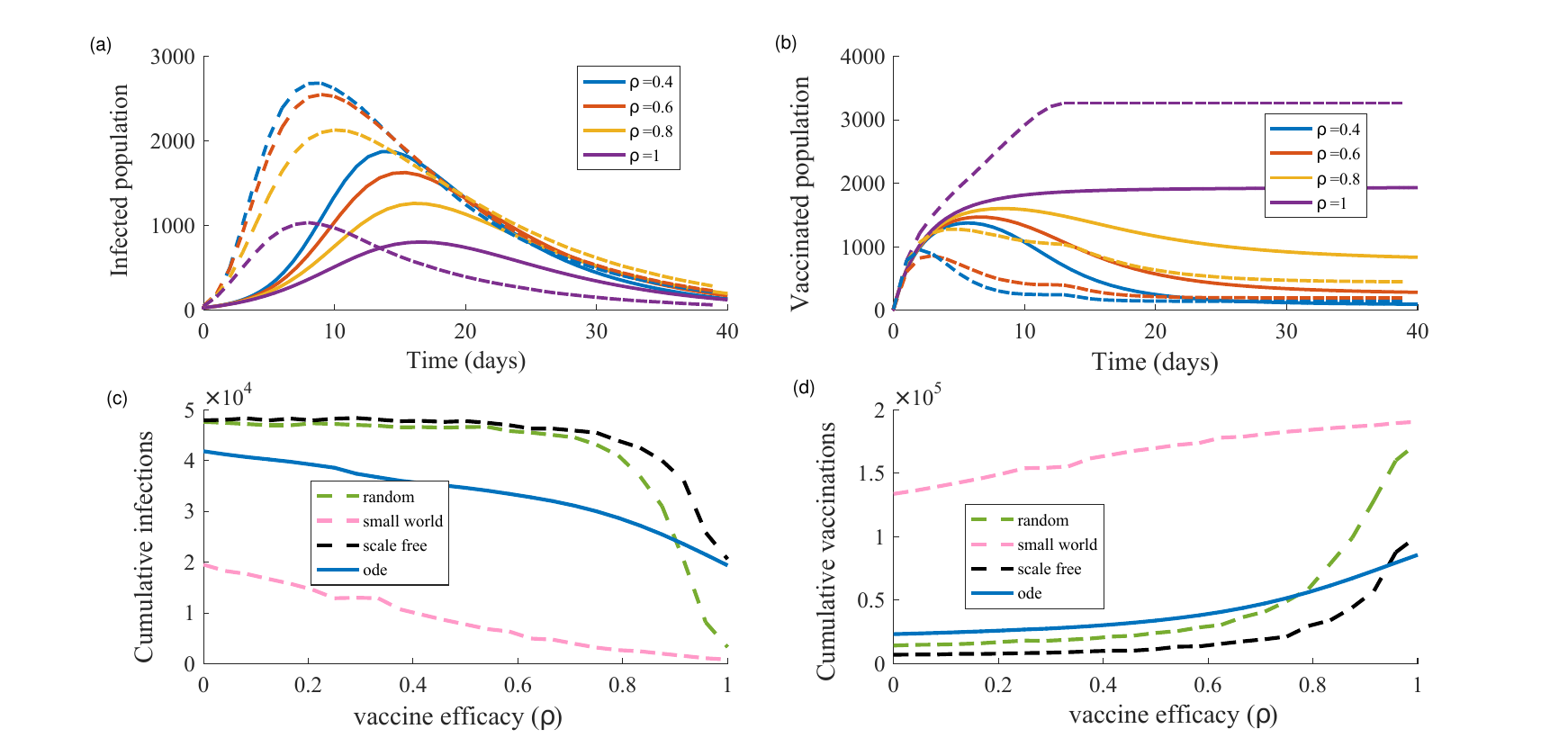}
	    \caption{Time series analysis of vaccine efficacy $\rho$ on (a) Infected population using the ODE model (solid lines) and scale-free network (dotted lines); (b) Vaccinated population using the ODE model (solid lines) and scale-free network (dotted lines). Figure (c) shows cumulative infections using the Ode model and different network topologies, and (d) shows the cumulative vaccinations using the ODE model and different network topologies as a function of different values of $\rho$. All the complex-network-based simulations are carried out by averaging 50 simulations.}
	    \label{f3}
	\end{figure}
 
 \subsubsection{Impact of vaccine efficacy and misinformation on disease dynamics across different network topologies}
Vaccine efficacy is crucial in changing the spread of infection by shaping the perceptions of individuals who become ill despite being immunized \cite{wang2024effect}. Inadequate vaccine effectiveness, directly and indirectly, impacts disease transmission \cite{sulayman2021sveire,rodrigues2020impact}. Figure \ref{f3} depicts the impact of vaccine efficacy on disease transmission across a range of scenarios, contrasting the outcomes from the ODE model and various network topologies. As vaccine efficacy rises, the peak of the infected population declines and occurs sooner in the scale-free network than in the ODE model (Fig. \ref{f3} (a)). The scale-free network indicates a higher peak and a slower decline, suggesting a longer persistence of the disease due to a small number of highly interconnected nodes (\textit{hubs}) that facilitate the spread of the disease (Fig. \textbf{S1}), making it more challenging to control even with higher vaccine efficacy (Fig. \ref{f3} (a)). Vaccine efficacy of 100 $\%$ results in a swift increase and a more significant number of vaccinated individuals (Fig. \ref{f3} (b)). This is because, in the scale-free network, the vaccination spreads more effectively after the highly connected hubs are vaccinated. Figure \ref{f3} (c) depicts the cumulative infections as a function of vaccine efficacy. In all scenarios, cumulative infections drop significantly as vaccine efficacy increases, with the ODE model showing a sharp decline, suggesting effective infection control.\\
Meanwhile, the random network predicts the highest cumulative infections among the network models, suggesting it is less effective at controlling disease spread than other network types. On the other hand, the small-world network predicts the lowest cumulative infections due to their high clustering and short average path lengths (Fig. \textbf{S4}), which enable rapid containment of the disease once key individuals are vaccinated, thereby reducing overall infections more effectively (Fig. \ref{f3} (c)). The cumulative vaccinations are relatively low across all network structures for low vaccine efficacy ($\rho < 0.4$) (Fig. \ref{f3} (d)). This suggests that when the vaccine is ineffective, the overall impact on the population remains limited. As $\rho$ increases beyond 0.4, cumulative vaccinations show a noticeable rise. This indicates that higher vaccine efficacy significantly enhances vaccination coverage. The ODE model consistently shows the highest cumulative vaccinations across the entire range of vaccine efficacy values due to the assumption of homogeneous mixing in the ODE model, leading to an overestimation of the vaccination spread compared to network-based models (Fig. \ref{f3} (d)). The scale-free network shows a steeper increase in cumulative vaccinations at higher vaccine efficacy due to highly connected hubs that can rapidly propagate the effects of vaccination (Fig. \ref{f3} (d)).\\

\begin{figure}[H]
	    \centering
	    \includegraphics[scale=0.6]{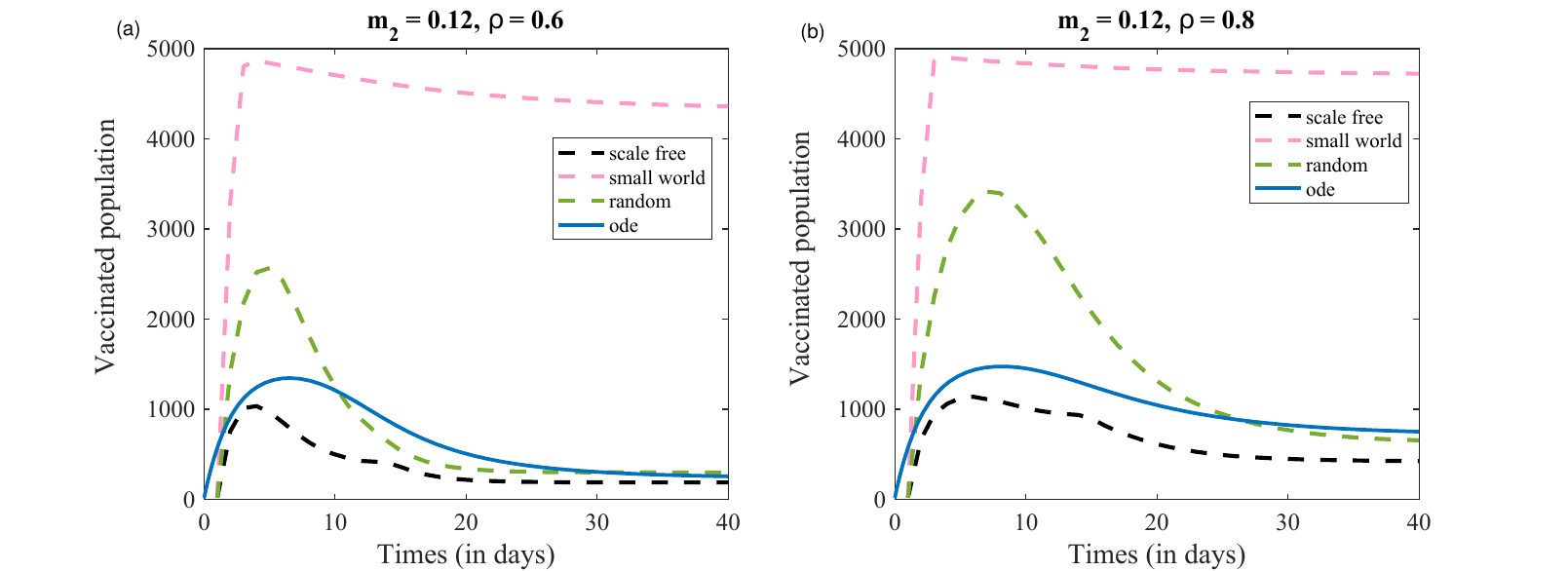}
	    \caption{Time series of the vaccinated population using different network structures for a fixed value of $m_2 = 0.12$ and varying vaccine efficacy. (a) Vaccination analysis with vaccine efficacy $\rho = 0.6$. (b) Vaccination time series with vaccine efficacy $\rho = 0.8$. As defined in Table \ref{table1}, other parameter values have remained unchanged. All complex network simulations are based on the average results from 50 runs.}
	    \label{f4}
\end{figure}
 
\begin{figure}[H]
	    \centering
	    \includegraphics[scale=0.6]{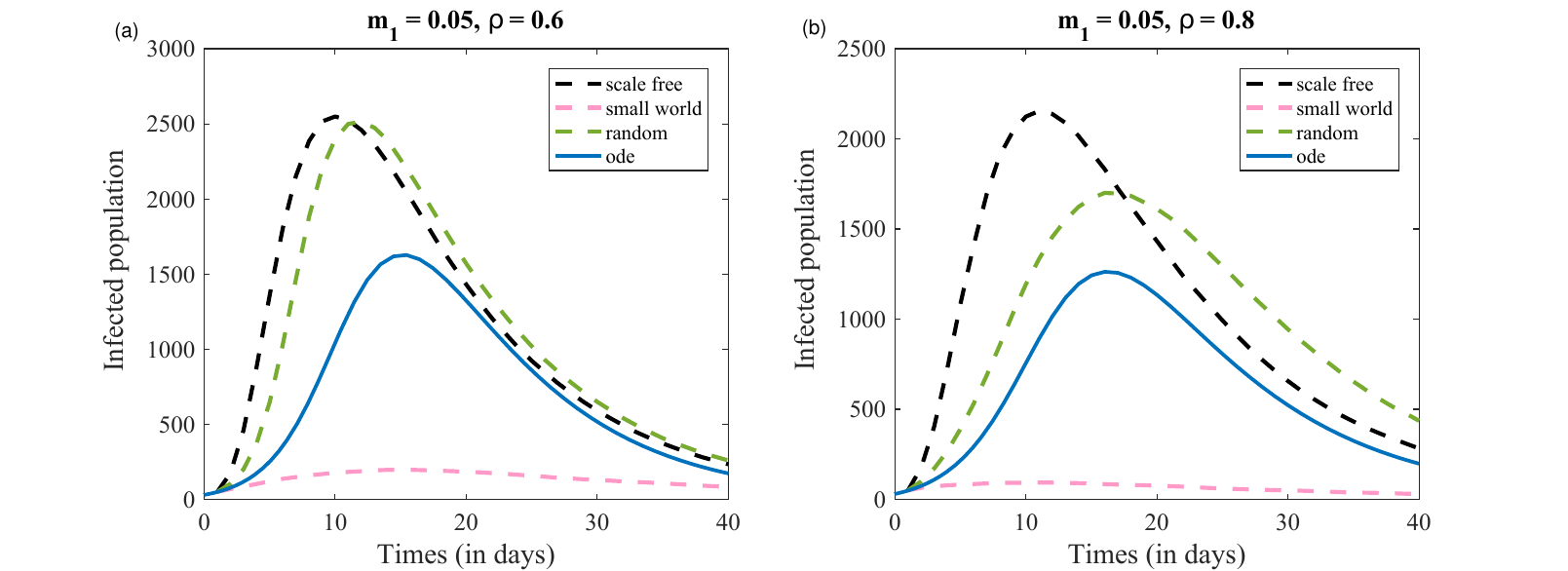}
	    \caption{ Figure shows time series of (a) infected population with vaccine efficacy $\rho = 0.6$. (b) infected population for vaccine efficacy $\rho = 0.8$; for fixed values of $m_1 = 0.05$, and other parameters will remains unchanged as defined in Table \ref{table1}. All complex network simulations are based on the average results from 50 runs.}
	    \label{f5}
\end{figure}

Misinformation about vaccines can lead to hesitancy, reducing their real-world impact even if they're highly effective \cite{lee2022misinformation,garett2021online}. While better vaccines can help counteract this somewhat, widespread misinformation can still significantly hinder vaccination efforts. Understanding this relationship is crucial for creating successful public health strategies. Figure \ref{f4} and \ref{f5} depicts the time series under various network structures and vaccine efficacy levels, with a fixed misinformation amplification parameter $m_2$ $\&$ $m_1$, respectively. With vaccine efficacy $\rho = 0.6$ (Fig. \ref{f4} (a)), the small-world network demonstrates rapid initial vaccination uptake, peaking early and then declining steadily due to the network's high clustering (Fig. \textbf{S4}), which facilitates the quick dissemination of information, including misinformation. This rapid decline occurs as misinformation spreads efficiently in a tightly-knit network, reducing the vaccination rate. The random network exhibits a moderate peak and gradual decline, reflecting its uniform degree distribution, which balances the spread of vaccination behaviour and misinformation. While higher vaccine efficacy, $\rho = 0.8$ results in a large vaccinated population for a random community population (Fig. \ref{f4} (b))). The scale-free network presents a moderate rise in vaccinated individuals, while the ODE model shows a smoother and lower peak, with a gradual decline, lacking the rapid spread dynamics of real-world networks. This indicates a more stable but less dynamic vaccination trend. Higher vaccine efficacy plays a crucial role in controlling the misinformation from the $I_V$ state individuals within the scale-free and random communities people (Fig. \ref{f4} (a \& b))).
Also, $m_1$ directly influences disease prevalence. Figure \ref{f5} shows the infection time series for fixed $m_1$ and varying $\rho$. For $\rho = 0.6$, the infection peaks in both scale-free and random networks are nearly identical due to higher reinfection rates (Fig. \ref{f5}(a)). For $\rho = 0.8$, the infected population is significantly higher in the scale-free network compared to the random network (Fig. \ref{f5}(b)), as high vaccine efficacy mainly affects individual connections, leading to more reinfections. Additionally, cumulative infections are greater for $\rho = 0.6$ than for $\rho = 0.8$ (Fig. \ref{f5}). In the ODE model, vaccine efficacy has no impact, as a homogeneous population does not influence reinfection rates.\\

Higher vaccine efficacy reduces infection peaks and accelerates disease control, with varying effectiveness across different network structures. Due to its hub-driven dynamics, the scale-free network shows higher cumulative infections than random and small-world networks. Misinformation from infected individuals (either $I_S$ or $I_V$) significantly impacts vaccination efficacy, particularly in networks with high clustering. Understanding these dynamics aids in developing targeted vaccination strategies and addressing misinformation, which is crucial for enhancing public health initiatives and improving disease control outcomes.

\subsubsection{Implications of disease transmission rate ($\beta$) and vaccine efficacy ($\rho$)}
\begin{figure}[H]
	\centering
   \includegraphics[scale=0.8]{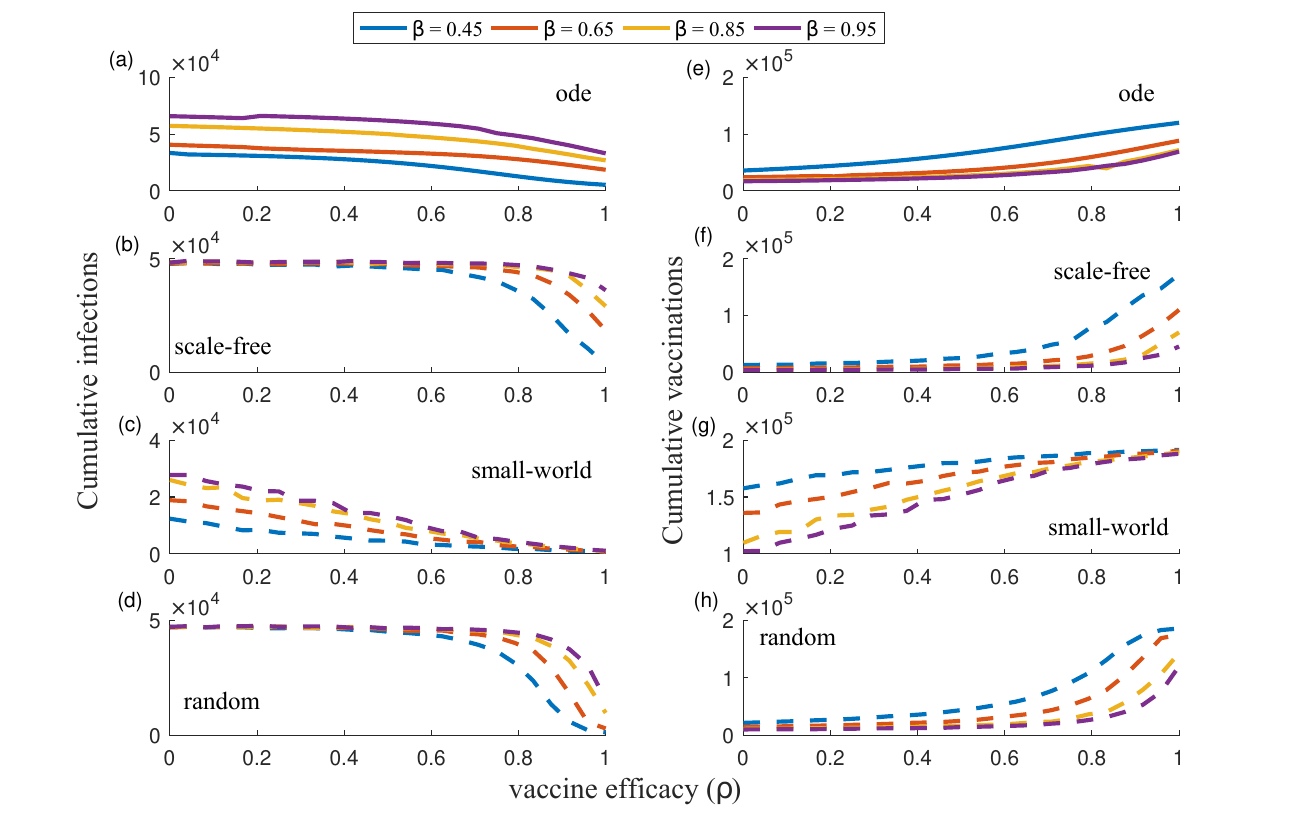}
	    \caption{Figure (a-d) represents the cumulative infected population for the ODE model, scale-free network, small-world network, and random network, respectively. (e-h) represents the cumulative vaccinated population for the ODE model, scale-free network, small-world network,
and random network, respectively, as a function of vaccine efficacy $(\rho)$ for different transmission rates ($\beta$). All complex network simulations are conducted by averaging the results from 50 separate runs.}
	    \label{fig:vwdiffrate}
\end{figure}

\begin{figure}[H]
	\centering
   \includegraphics[scale=0.45]{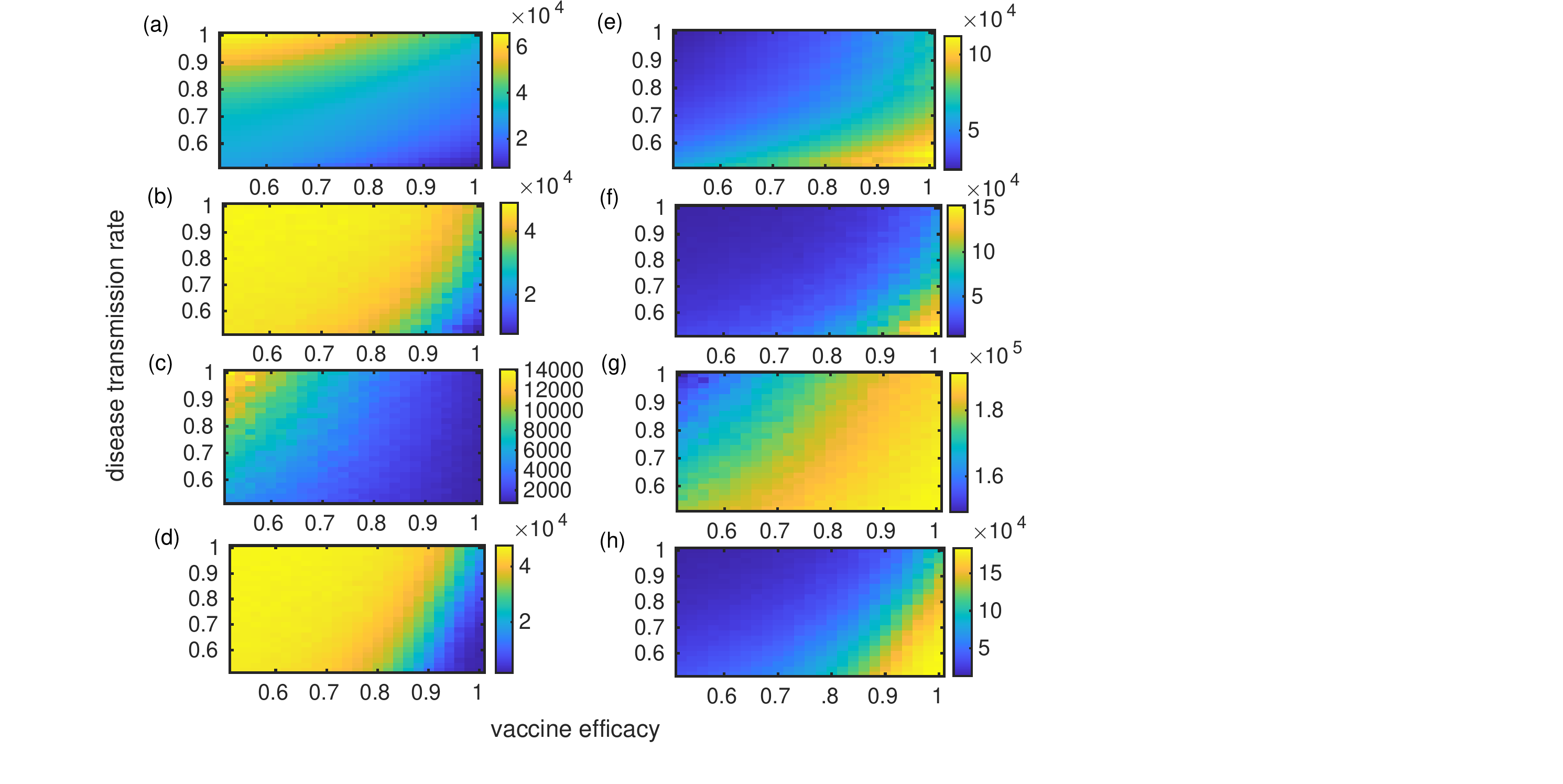}
	    \caption{Figure (a-d) represents the cumulative infected population for the ODE model, scale-free network, small-world network, and random network, respectively. (e-h) represents the cumulative vaccinated population for the ODE model, scale-free network, small-world network, and random network, respectively, as a function of vaccine efficacy $(\rho)$ on the x-axis and disease transmission rate ($\beta$) on the y-axis. All the points in Figure (b-d) and (f-h) are the average of 25 simulations.}
	    \label{fig:surfbeffi}
\end{figure}

Higher vaccine effectiveness is associated with a significant reduction in overall disease prevalence \cite{halloran1991direct}. Effective vaccines reduce disease incidence and protect non-vaccinated individuals \cite{halloran2010evaluating}. Also individual contact structure plays a pivotal role in the spread of infectious diseases. Network-based models provide insights into how individual interactions influence transmission pathways and how targeted vaccination strategies can mitigate disease spread \cite{kao2010networks}.\\ Figure \ref{fig:vwdiffrate} demonstrates the impact of vaccine efficacy on disease transmission rates across various scenarios. Figure \ref{fig:vwdiffrate} (e-h) shows that cumulative vaccination is higher for low disease transmission rate ($\beta)$, regardless of the percentage of vaccine effectiveness ($\rho)$.
Consequently, higher $\beta$ values adversely affect disease propagation (Fig. \ref{fig:vwdiffrate}). Our model simulations across different networks and the homogenous population represent that in different communities, the cumulative infection is lower across all networks compared to the ODE model (Fig. \ref{fig:vwdiffrate} (a) \& (b-d)). For lower vaccine efficacy ($\rho \leq 0.5$), cumulative infections are almost identical in scale-free and random networks. This similarity is due to the uniformity in population characteristics. In homogeneous populations, the infection spreads more uniformly, making the specific network topology less critical in determining cumulative infection. As vaccine efficacy increases, the cumulative vaccination rate within the population (whether homogeneous or heterogeneous) also increases for any disease transmission rate.
As vaccine efficacy may alter the reinfection pattern and disease transmission rate can boost the prevalence, we simulated our model for varying vaccine effectiveness under different disease transmission rates. Figure \ref{fig:surfbeffi} summarizes these critical parameters' dynamics. Higher vaccine efficacy leads to increased vaccine coverage (Fig. \ref{fig:surfbeffi} (e-h)) and decreased infection rates (Fig. \ref{fig:surfbeffi} (a-d)). Notably, the transition from low to high coverage or high to low prevalence occurs at lower transmission rates as vaccine efficacy increases. For instance, at a vaccine efficacy of $\rho = 0.5$, this transition occurs at $\beta = 0.65$, whereas at $\rho = 1$, it occurs at $\beta = 0.55$. In scale-free networks, cumulative prevalence is significantly higher than in other network topologies and the ODE model (Fig. \ref{fig:surfbeffi} b \& (a,c,d)) due to highly connected hubs. As $\rho$ increases, reinfection rates decline, leading to fewer infections and more people opting for vaccination (Fig. \ref{fig:surfbeffi} (e-h)). 
Interestingly, the small-world network topology shows higher cumulative vaccination coverage compared to other topologies (Fig. \ref{fig:surfbeffi} g \& (e, f, h)) due to their path length. While the infected neighbours in the ODE model do not influence outcomes due to the homogeneous population, demonstrating similar patterns for higher vaccine efficacy and low transmission rates (Fig. \ref{fig:surfbeffi} a \& e).\\
Our study demonstrates the critical role of vaccine efficacy in reducing disease transmission and reinfection rates, which lead to a jump in misinformation. Additionally, the impact of network structures on disease spread and vaccination coverage suggests that public health policies should implement targeted vaccination strategies, particularly in densely connected populations, to maximize coverage and minimize transmission. Higher cumulative vaccination coverage observed in small-world networks indicates that public health interventions can especially be effective in similar populations, guiding the design of efficient vaccination campaigns to reach the most vulnerable and interconnected individuals swiftly.
\subsubsection{Influence of vaccination rates and misinformation on disease prevalence}
\begin{figure}[H]
	\centering
   \includegraphics[scale=0.45]{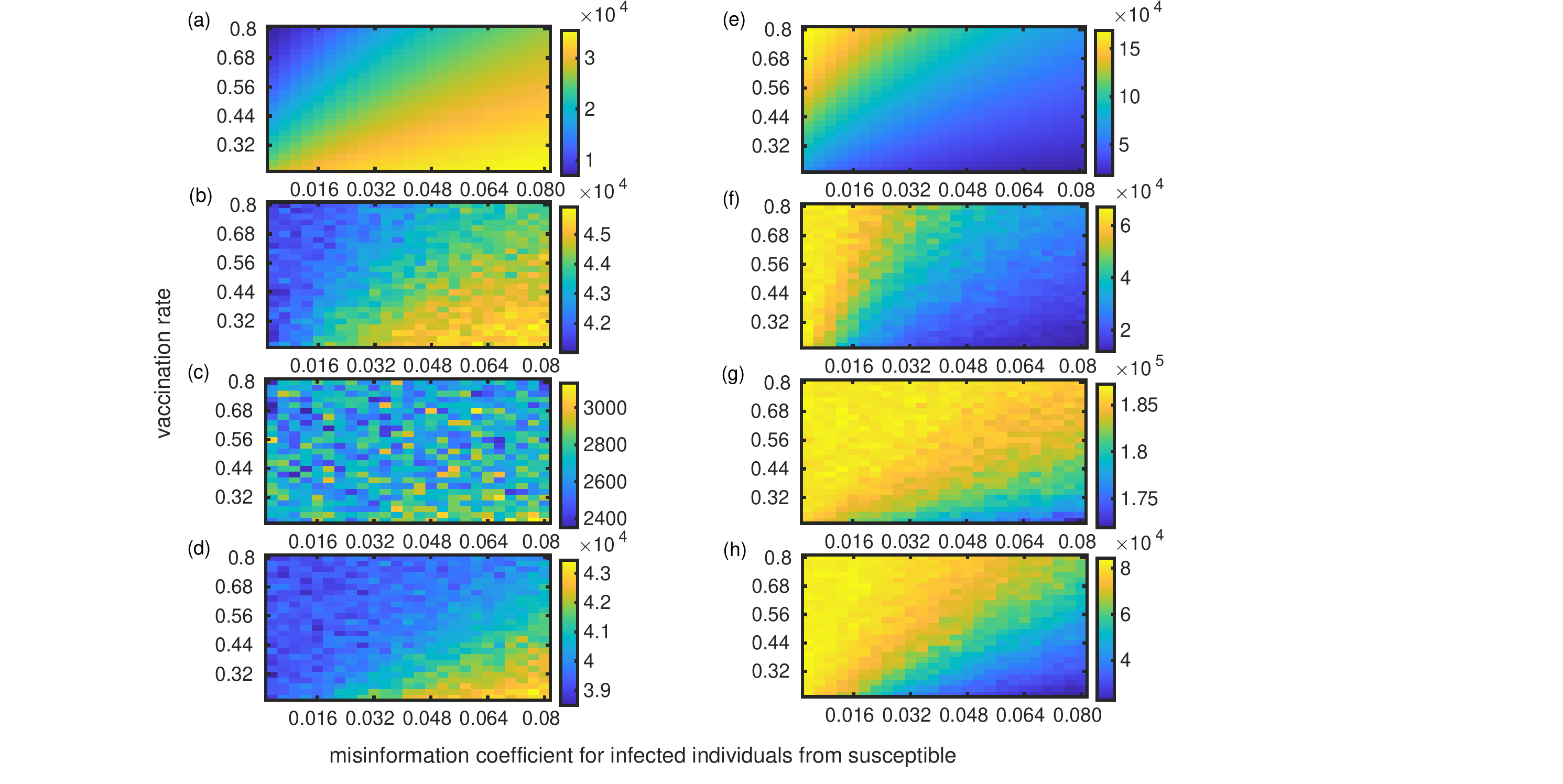}
	    \caption{Figure (a-d) represents the cumulative infected population for the ODE model, scale-free network, small-world network, and random network, respectively. (e-h) represents the cumulative vaccinated population for the ODE model, scale-free network, small-world network, and random network, respectively, as a function of the misinformation coefficient for infected individuals from the susceptible state ($I_s$) on the x-axis and vaccination rate ($v_1$) on the y-axis. All the points in Figure (b-d) and (f-h) are the average of 25 simulations.}
	    \label{fig:m1v1surf}
\end{figure}
Vaccination rates are critical in controlling disease outbreaks and achieving public health goals \cite{schumacher2021increasing,angeli2022modeling}. However, the effectiveness of vaccines can be undermined by misinformation, which plays a significant role in shaping public perception and vaccine hesitancy \cite{de2022fake,honora2022does}. Our results highlight that even highly effective vaccines can have a diminished impact when misinformation is prevalent, illustrating the need for effective strategies to counteract misinformation. Community structures respond differently to misinformation and vaccination efforts \cite{petrizzelli2022beyond}. Understanding these variations, we have simulated our model for various values of $m_1$ and vaccination rate ($v_1$) for different populations (Fig. \ref{fig:m1v1surf}). Figure \ref{fig:m1v1surf} illustrates the dynamics of cumulative infection and vaccination rates as a function of the misinformation amplification parameter ($m_1$) and vaccination rate ($v_1$) across the homogenous population and three network topologies. Across all topologies, an increase in $m_1$ leads to a rise in cumulative infections (Fig. \ref{fig:m1v1surf} (a-d)), indicating that higher misinformation significantly accelerates disease spread, especially when $v_1$ is low. In the ODE model, which represents a homogeneous population, the infection rate increases rapidly as $m_1$ rises, but high vaccination rates $v_1$ effectively reduce cumulative infections (Fig. \ref{fig:m1v1surf} (a)). In contrast, the Small-World network, which features local clustering, shows a more gradual infection spread. Here, high $v_1$ values create localized pockets of vaccinated individuals, slowing down infection, although low vaccination rates coupled with even moderate misinformation levels lead to higher cumulative infections (Fig. \ref{fig:m1v1surf} (c \& g)).\\

Meanwhile, The Erd\H{o}s-R\'enyi (random) network demonstrates that infection spreads uniformly. At the same time, high vaccination rates significantly reduce infection levels; the random nature of the network makes it highly susceptible to increased $m_1$, leading to faster disease spread under high misinformation (Fig. \ref{fig:m1v1surf} (d \& h)). Interestingly, the scale-free network displays the most sensitive behaviour to misinformation. At low $m_1$, the hubs act as vaccination strongholds, dramatically reducing cumulative infections. However, at higher $m_1$, if hubs remain non-vaccinated due to misinformation, they become super-spreaders, resulting in a sharp spike in cumulative infections (Fig. \ref{fig:m1v1surf} (b \& f)).\\ 

We can observe that the scale-free and small-world networks are particularly vulnerable to misinformation amplifying infection when nodes are affected. Comparatively, vaccination is more effective in the ODE and Erd\H{o}s-R\'enyi networks due to their uniform structure (Fig. \ref{fig:m1v1surf} (e \& h)), where higher $v_1$ consistently leads to a sharp decline in infection. Our results highlight the varying influence of network topology on disease dynamics, with homogeneous and random networks being more controllable through vaccination.\\

Our observations underscore the critical need to counteract misinformation, especially in populations with highly structured social networks. Targeted vaccination campaigns can drastically reduce disease spread, mainly aimed at influential individuals or hubs within these networks. Public health strategies should prioritize increasing overall vaccination rates and addressing misinformation in critical groups, thus improving the effectiveness of vaccination efforts in mitigating outbreaks.
\section{Discussion}
Social networking sites have become the epicentre of vaccine-related misinformation, which raises significant concerns for public health. The incessant viral sharing of false or misleading information on these platforms can foster vaccine hesitancy, leading to declining immunization rates. For instance, during the COVID-19 pandemic, anti-vaccine messages proliferated on various platforms, causing some people to doubt the safety of vaccines. Similarly, in cases like measles outbreaks, fears about vaccines are often unfounded but are amplified by online claims that vaccinations cause serious diseases, which then lead to reduced vaccination rates. Social media spreads false information rapidly due to its viral nature, causing misinformation to travel faster and farther than the truth. This leads people to base health decisions on incorrect or misleading materials, undermining disease prevention efforts. This cycle of misinformation and changing behaviour affects disease dynamics, creating pockets of unvaccinated populations and increasing the risk of new outbreaks that could otherwise be avoided.

The impact of vaccine hesitancy due to misinformation can be better understood by analyzing its effects on disease dynamics. This involves modelling how reduced immunization levels impact the spread of infectious diseases, making populations more susceptible to more significant outbreaks, prolonged epidemics, and increased morbidity and mortality. To understand the effects of declining vaccination rates and reduce health risks from misinformation, researchers must include social behaviour factors, such as misinformation spread, in disease transmission models. This approach is essential for developing effective strategies. Understanding these dynamics is key to designing effective interventions that restore vaccine confidence and control the spread of infectious diseases.

With these concerns in mind, we propose a mathematical model incorporating the effects of ``misinformation". Here, we analyzed how vaccination uptake is affected by misinformation and its impact on both homogeneous and heterogeneous populations. Various computational and mathematical models have been used to examine how misinformation affects vaccination campaigns in different populations \cite{tambuscio2015fact,zhang2016dynamics,loomba2021measuring,prieto2021vaccination,pierri2022online,mumtaz2022exploring,chen2022coevolving,sun2023finding}. However many of these studies do not provide a direct comparison to explore how misinformation influences homogeneous versus heterogeneous populations. Moreover, the interplay between population heterogeneity, the rate of misinformation, and vaccination rates across different network structures has not been thoroughly investigated. This research is essential when designing successful vaccination strategies, particularly during pandemics. Understanding how misinformation impacts vaccination rates in different populations and network structures is essential for developing effective public health policies.

In our study, we used numerical simulations with different network approaches ({\textit{scale-free networks, random networks, and small-world networks}}) and ordinary differential equations (ODEs) approaches. In Figure \ref{f3}, we first evaluated the impact of vaccine efficacy on infected and vaccinated individuals. It is clear that increasing vaccine efficacy consistently reduces infections in all models. However, the network structure greatly influences the outcome. Infections rise and fall more drastically in the scale-free network due to its high-degree nodes, whereas the small-world network exhibits the best performance in controlling disease spread, as indicated by lower cumulative infections. The ODE model tends to overestimate vaccination coverage compared to network models, especially when vaccine efficacy is high. This underscores the need to consider population structure when evaluating the impact of vaccines.

Furthermore, as depicted in Figure \ref{f4}, with a constant misinformation factor $m_2$, we observe the influence of different network structures and vaccine effectiveness on the time-varying vaccinated population. Vaccination levels peak early in small-world networks and then sharply decline due to the close-knit nature of neighbourhoods and information flow. Random networks exhibit a more gradual peak and dip, while scale-free networks show a slower increase in vaccination uptake. The ODE model, on the other hand, presents a continuous line without the abrupt changes seen in actual networks.  Notably, the population vaccinated is more significant due to the higher vaccine efficacy $(\rho = 0.8)$, especially in random and scale-free networks.

Next, we analyzed the impact of vaccine efficacy with a fixed misinformation factor $(m_1)$. Figure \ref{f5} shows that infection dynamics change based on different levels of vaccine efficacy. At $\rho = 0.6$, infection peaks are similar in scale-free and random networks. However, at $\rho = 0.8$, scale-free networks have higher infection rates due to the possibility of reinfections occurring within hubs. High vaccine efficacy reduces peak infections and total cases across all models.

Next, we assessed the effect of vaccine efficacy on disease transmission across various scenarios. As shown in Figure \ref{fig:vwdiffrate}, cumulative vaccinations are high when the disease transmission rate $(\beta)$ is low, regardless of vaccine efficacy $(\rho)$. However, higher transmission rates pose challenges for disease control. The ODE model typically shows higher infection levels than network-based models. For vaccine efficacy $\left(\rho \leq 0.5\right)$, cumulative infections are similar in scale-free and random networks. Increasing vaccine efficacy consistently boosts cumulative vaccination coverage, regardless of transmission rates or population structure.

Additionally, we examined the role of vaccine efficacy and disease transmission rates in reinfection prevalence and disease causation. Figure \ref{fig:surfbeffi} demonstrates that higher vaccine efficacy leads to increased vaccination coverage and lower infection rates. The transition from low to high vaccination coverage occurs at lower transmission rates when vaccine efficacy is high. For example, at $\rho = 0.5$, the transmission rate is 0.65; at $\rho = 1$, it shifts to 0.55. Scale-free networks have higher cumulative infection rates due to their dense hubs, while smaller communities achieve higher immunity levels. These results underscore the importance of specialized vaccination strategies, particularly in highly interconnected populations.

Furthermore, we investigated the impact of misinformation and vaccine uptake on disease transmission across various network types. Figure \ref{fig:m1v1surf} shows that higher levels of misinformation $(m_1)$ correlate with increased infections, especially at lower vaccination rates $(v_1)$. In the ODE model, which represents a uniform society, higher vaccination rates reduce infections, even in the presence of misinformation. However, this is not the case in scale-free and small-world networks, where misinformation exacerbates disease spread. In these networks, hubs act as super-spreaders at high misinformation levels $(m_1)$. On the other hand, random networks show more uniform infection patterns but still allow for the spread of fake news. Therefore, targeted vaccination campaigns through organized networks are essential to mitigate the effects of misinformation and control disease transmission.

In summary, our study demonstrates the significant influence that vaccine efficacy, misinformation, and network structures have on disease spread. Our simulations show that improving vaccine efficacy reduces infection rates and increases vaccination coverage across different network structures. However, misinformation undermines these efforts, particularly in large, connected networks like scale-free and small-world networks, where key individuals significantly spread disease. These findings highlight the need for public health interventions that consider both population structures and the detrimental effects of misinformation. Targeted vaccination programs focusing on influential individuals or "hubs" in highly connected networks can significantly reduce disease transmission and counteract the negative impacts of misinformation.
Furthermore, sustained information campaigns to counter misinformation are essential to maintain high vaccination rates, particularly in vulnerable groups.\\

Our research provides important insights into how vaccination decisions, network evolution, and disease transmission affect communities. We quantify the impact of misinformation on vaccination rates in both homogeneous and heterogeneous populations. Our results show that misinformation significantly lowers vaccination rates, particularly in homogeneous groups, while heterogeneous populations are more resilient. Small-world networks tend to achieve higher vaccination rates, regardless of vaccine effectiveness, while scale-free networks experience lower coverage as misinformation increases. Notably, cumulative infections do not depend on disease transmission rates when vaccine efficacy is only partial. In small-world networks, the number of infections varies significantly across different vaccination rates and misinformation levels, whereas vaccination rates are highest when misinformation is low. Public health efforts should focus on combating misinformation to control disease spread, especially in homogeneous populations and scale-free networks where its effects are most potent. Strengthening community networks and providing accurate vaccine information can help to improve vaccination rates. Even when vaccine efficacy differs, concentrating public health campaigns on small-world networks can enhance vaccine uptake. These insights can guide public health policymakers in creating effective vaccination strategies considering population diversity, ultimately helping in reduction of community infections.\\

Current research offers insights into the relationships between misinformation, vaccination rates, and their impact on disease prevalence. However, we must recognize the limitations of our model and the opportunities for enhancement by integrating more realistic scenarios. The associated risks are often uncertain in real-life situations, especially with newly introduced vaccines. This uncertainty can lead to the spread of rumours on social media about potential adverse effects, such as paralysis, lung congestion, or even death. Such adverse events can significantly shape the perceived risks of vaccines within a community \cite{bhattacharyya2019impact,mushtaq2022review}. This concern is also relevant to emerging pathogens like COVID-19.
Another factor to consider is the daily evolution of social networks. During a pandemic, individuals limit physical contact with sick relatives and friends and often practice social isolation. This behaviour leads to changes in social networks. Our model could be improved by incorporating these limitations, highlighting new scope of future research.\\

Our research offers a promising framework for creating more effective and customized strategies to manage vaccine-preventable diseases by examining the interactions between misinformation management, awareness promotion, homogeneous populations, and network structures. In our highly interconnected world, improving disease control strategies and advancing public health initiatives can be achieved by utilizing social media platforms to share accurate information about vaccines and infections.

\section*{Acknowledgements:}  The research work of Jai Prakash Tripathi is supported by the Science and Engineering Research Board (SERB), India [File No. MTR/2022/001028].

\bibliographystyle{unsrt}
\bibliography{ref}
\end{document}